\documentclass[5p]{elsarticle}
 \usepackage{amssymb}
\usepackage{lipsum}
\journal{Physics Letters B}

\begin{document}
\begin{frontmatter}   
\title{Neutron Stars  constraints on a late G transition}
 \author[]{Itzhak Goldman$^{a, b}$ }
 
 \affiliation[first]{organization={Physics Department Afeka College},
              addressline={}, 
              city={Tel Aviv},
            postcode={6998812}, 
                       country={Israel}}
            
         \affiliation[]{organization={Astrophysics Department Tel Aviv University},
            addressline={},
            city={Tel Aviv},
            postcode={6997801}, 
                      country={Israel}}

  \begin{abstract}
  It has been  suggested recently that the Hubble tension could be eliminated by a sharp, $\sim 10\%$ increase of the effective gravitational constant at  $z \sim 0.01$. This would  decrease the luminosities of type 1a supernovae in just the needed amount to explain the larger value of the Hubble parameter. In the present paper we call attention to a dramatic effect  of such transition on neutron stars. A neutron star that existed at $z=0.01$  would contract, conserving the baryon mass but  undergoing  a mass reduction. We  computed  neutron star models, with a realistic equation of state, and obtained  that this reduction is  typically $ 0.04 M_{\odot}$. This amounts to an energy  of $7 \times 10^{52}$ erg. The transition  will affect {\it all} neutron stars that   formed along the history of each galaxy prior to the transition.  Given the large number of  neutron stars per galaxy,   the liberated energy  is huge. An estimate of the expected fluxes of neutrinos and x-rays  yields values exceeding observational upper limits, thus rendering the  late G transition scenario non-viable.
  \end{abstract}
  

\end{frontmatter}

\section{Introduction}

The Hubble tension refers to  the   disagreement
between the value of $H_0$  inferred from Planck observations of the CMB \citep{Planck20}:  $(67.4  \pm 0.5$ km/(s Mpc) and the value derived  from type 1a  supernovae data, e.g. by the SH0ES collaboration \citep{Riess+21}: $(73.0 \pm  1.4)$ km/(s Mpc).

Recently, it has been   suggested  \cite{Marra+Perivolaropoulo21}
   to resolve the Hubble tension by  postulating  a sharp increase   of the effective gravitational constant by $10 \%$ at 
a redshift of $\ 0.01$. The increase in the value of G would have yielded smaller values of the   type 1a supernovae luminosities and therefore  a higher  $H_0$.  In addition these authors note that such a transition could alleviate the growth tension. The late G transition scenario  has been further discussed  by  \cite{Alestas+Perivolaropoulos21, Alestas+21, Alestas+22a, Alestas+22b}.

On the other  hand \cite{Alvey+20}  examined  primordial nucleosynthesis and found that the value of  the effective gravitational constant at that era  was the same as that of the present epoch. Also, 
  \cite{Sakr+Sapone22} argued that CMB data  disfavor a change of G.

The present work focuses on   implications of a sharp late G transition on neutron stars.  
The increase of G would   induce  a shrinking of the neutron star and heating it. After the release of the thermal energy,  the star will settle to a new hydrostatic equilibrium, with   an unchanged baryon number but an increased binding energy,  and smaller mass. This will occur for all neutron stars that formed prior to the transition.

In section 2.  static neutron star models before and after the transition (with the same baryon mass) are constructed in order to estimate the mass reduction. In section  3.  
the observational implications of the released  energy are examined. In section 4. discussion and conclusions are presented.

  \section{Neutron star models}
  We computed static non-rotating neutron star models employing a realistic - based on observations- nuclear matter equation of state
  \citep{Steiner+10}. The goal is to obtain a typical representative value of the expected mass reduction.
 The spherical symmetric static line element used is

\begin{equation}
ds^2= e ^{2 \phi(r)}c^2 dt^2 - e^{2\lambda(r)}dr^2 - r^2  d\Omega^2
\end{equation}

  The equation of state determines the pressure $p$ as function of the  energy density 
  $\rho$.
  The   structure TOV equations   are  \citep{Tolman1939, Oppenheimer+Volkoff1939}  
   
 \begin{eqnarray}
m'(r)= 4\pi r^2 \rho(r)\\
e^{2\lambda(r)} =\left(1- \frac{2 G m(r)}{c^2 r}\right)^{-1}\\
p'(r)= -\frac{G \left(\rho(r) + p(r)\right)  }{ c^2 r^2 \left(1- \frac{2 G m(r)}{c^2 r}\right)} \left(m(r)+ 4 \pi p r^3  c^{-2} \right)\\
\phi '(r)= -\frac{p'(r)}{\rho(r) + p(r)}
 \end{eqnarray}

$m(r)$ is the mass enclosed within a radius $r$, the star total mass is $M= m(R)$, where $R$ is the star radius, defined by the vanishing of the pressure $p(R)=0$.  
  
  The baryon number density $n(r)$  corresponding to a given energy density $\rho(r)$
is obtained   from the relation
  \begin{equation}
  \frac{n'(r)}{n(r)}= \frac{\rho '(r)}{\rho(r) + p(r)}
  \end{equation}
  where a prime  denotes   derivative with respect to $r$.The total baryon mass of the star, $M_b$,  is given by 
  \begin{equation}
 \hspace{-1cm} M_b =m_b\int_0^R \sqrt{-g}n(r) u^0  d^3x= m_b\int _0^R 4\pi r^2 e^{\lambda(r)}n(r)dr
\end{equation}

with    $ \sqrt{-g}= e^{(\lambda(r) + \phi(r))}4\pi r^2$ and $e^{\phi} u^0=1$. Employing equation (3) yields
  \begin{equation}
  M_b  = m_b  4\pi\int_0^R n(r)\left(1- \frac{2 G m(r)}{c^2 r}\right)^{-1/2}r^2 dr
  \end{equation}
  
  For a given energy density at $ r=0$ the structure equations are solved and yield a neutron star model.
  Then, the equations with the increased gravitational constant are solved. The center energy density is varied until the baryon mass is the same as that in the pre-transition model.
  
 Table 1 displays four such pairs of neutron star models,  before and after the G increase. Each pair corresponds to a given total baryon mass.  The 
  post transition models have a smaller radius, a smaller mass and  larger surface redshift. The mass reduction and the increase in the surface redshift are  larger  the larger the total baryon mass. Based on the computed models, we adopt as a representative  value for the mass reduction:     $\Delta M=0.04 M_{\odot}$  corresponding to energy release 
 of $7 \times 10^{52}$erg, namely about $20\ \%$ of that released in a typical  core collapse supernova.
 
	\begin{table} 
	\centering
	\caption{Neutron stars parameters before and after G transition. $M_b$ is the baryon mass, $M$ is the mass, $R$ is the radius and$z_s$  is the surface redshift    before the transition. The corresponding parameters after the transition are denoted by a prime. Masses are in units of solar mass ,$M_{\odot}$, and radii in units of $ km$.}
   
      \begin{tabular}{c c c c c c c} 
   $ M_b $ & $M$&$R$  & $z_s$ &$M'$&$R'$ & $z_s'$  \\
   \hline
   \hline
  1.60 & 1.41&10.8& 0.28 & 1.38 &10.43 &0.33\\ 

1.77 &1.54 & 10.9 &0.31 &1.51 &10.46&0.37\\ 
 
  2.06& 1.75&10.97 &0.38 & 1.71 & 10.27 &0.48\\ 
 
   2.20 &1 .86&10.92&0.42 &1.81 & 9,63 &0.60 \\ 
  \hline
     \end{tabular} 
 \end{table}
   \section{Observational implications}
   The transition will affect {\it all} neutron stars   formed along the history of each galaxy, prior to the transition.   
     The authors of  \cite{Sartore+2010}, assuming a constant rate of formation over time, estimated the total number of neutron stars in the Milky Way Galaxy (MW) to be $10^8- 10^9 $.  To estimate the number of neutron stars in the Galaxy that formed at redshifts larger than $0.01$ we note that  the age of the Galaxy ($\sim 13.6~ Gyr$) is 2 orders of magnitude larger than the co-moving time since $z=0.01$ (see section  3.1). Thus, we adopt a representative value of $ 5\times 10^8$ for the number of neutron stars  in a Milky Way type galaxy that were formed  before the transition. 
     
    Therefore, for each  Milky Way type galaxy the total energy  released in the transition is
 $\sim 3.5 \times 10^{61}$~erg. This is indeed a dramatic event.In the following we examine the implications regarding the viability of the proposed prompt late G transition.  
 
 \subsection{X-ray flux from the Milky Way galaxy neutron stars}
 
  First, we check wether  there is an observable constraint from the MW neutron stars. To find out one should obtain the time elapsed, $t_{tr}$, from the transition to the present time. The relevant relations are, assuming  a spatially flat cosmological model,

\begin{eqnarray}
t_{tr} = H_0^{-1}\int_0^{0.01} E(z) (1+z)^{-1}dz\\
E(z)= H(z)/H_0= \left(\Omega_m(1+z)^3+ \Omega_{\lambda}\right)^{\frac{1}{2}}
\end{eqnarray} 
 
 For $H_0=73 Mpc/s/km$,   $\Omega_m=0.3$, and $\Omega_{\lambda}
  =0.7$, the resulting value is
 $t_{tr}= 1.36\times 10^8$~yr. 
 
The authors of  \cite{Sartore+2010} concluded that most of the neutron stars in the MW galaxy are located between distances of $4$ to $6$ Kpc from the galactic center.
           
  Given the solar system distance of  $8$ Kpc from the galactic center,   neutrinos and photons         
 arriving at the present time to earth were emitted  from neutron stars about $\sim 10^4$yr after the transition,  from the  most distant neutron stars, and  $\sim 10^5$yr  from the closest ones.  By these times there will be no neutrino emission. There will be however photon emissions from the surface of the neutron stars. According to \citep {Yakovlev+2001,Yakovlev+Pethick04,Potekhin+20} for non-superfluid neutron stars with nucleonic core   the  surface temperature of the neutron stars (as measured by  a distant observer) after $10^4 $yr  would be  be  $T_{s\infty}\sim 10^6$~K with luminosities observed far from each neutron star $L_{\infty}\sim 10^{34} erg\  s ^{-1}$. Taking the number of the  most distant  galactic neutron stars to be $1\times 10^8$ yields a total luminosity of $1\times  10^{42} erg \ s^{-1}$ of $\sim 0.1$KeV X-rays. The flux from these neutron stars will be 
 \begin{equation}
 F_x=   \frac{10^8 \times 10^{34}}{ 4 \pi  (14\times 3.1 \times 10^{21})^2}=4. \times 10^{-5}\  erg \ s^{-1}\ cm^{-2}
\end{equation}
 The unresolved, observed diffuse X-ray background flux in the range of $0.5 - 1$~ KeV obtained  by   the Chandra COSMOS Legacy Survey \cite{Cappelluti+17} is 
\begin{equation}
 (1.24\pm 0.17  )\times 10^{-12} {\rm \  erg \ s^{-1}\ cm^{-2}\deg^{-2}} 
  \end{equation}
  The imaged field was $ 2.2 \deg^2$ implying a flux  of 
  \begin{equation}
  (2.73\pm 0.37)   \times 10^{-12}\  erg \ s^{-1}\ cm^{-2} 
  \end{equation}
which is 7 order of magnitude smaller than the G transition prediction.

 \subsection{neutrino flux from external galaxies}
Next, we consider galaxies whose photons and neutrinos arrive today at Earth.  The luminosity distance to these galaxies is calculated using
 
 \begin{equation}
 d(z)= c  H_0^{-1}(1+z)\int_0^z\frac{1}{E(x)} dx
 \end{equation}
For the parameters used above, the  luminosity distance is $d=d(0.01)=41.5\ Mpc$.

 Following a sharp increase of the effective gravitational constant,  a neutron star will settle in its new state in a hydrodynamical time scale which is of the order of a millisecond, given that the sound speed is a fraction of the speed of light.  The released energy will be then trapped as thermal energy. Most of the  energy will be emitted by  $\sim 10$~ MeV neutrinos over a time of  few tens of seconds. Each such  neutrino burst will carry
    a total energy of $ 7 \times 10^{52} erg$. However, the total neutrino energy flux from such  a galaxy will be spread over a   time span that depends on the galaxy     inclination and the spatial distribution of neutron stars within it. We adopt the  distribution suggested by \cite{Sartore+2010} also  for external MW type galaxies.    
     
 For  a number density of galaxies at $z\sim 0.01$ of $0.1\ Mpc ^{-3}$ \citep{Conselice+2016},  the 
    number of galaxies in  a spherical shell surrounding the observer with radius of $41.5$~Mpc and thickness of $50$~Kpc is  $\sim 100 $. 
The  longest time spread is for an edge-on galaxy and would be $\sim 3.9 \times 10^4$yr and consequently the minimal luminosity   from a galaxy would be    $2.84\times10^{49} erg\ s^{-1}$,  and the minimal flux from a galaxy would be $1.4\times 10^{-4} erg\  s^{-1} \ cm^{-2}$.
   Thus, the flux from all the $\sim 100$ galaxies  would be  $\geq 1.4\times 10^{-2} {\rm erg\  s^{-1} \ cm^{-2}}$. 
    
The  Super-Kamiokande observational upper limit on diffuse neutrino background at neutrino energies of   9.3-11.3 MeV
is $5.9 \times 10^{-4} erg\  s^{-1}\ cm^{-2}$\citep{Abe+2021} in conflict with the minimal  expected value according to the G transition scenario, estimated above.

     \subsection{X-ray flux from external galaxies}
In the case of a prompt G transition, each neutron star will emit most of the deposited thermal energy  via  neutrinos. The luminosity of thermal photons emitted  from the surface,  measured by a distant observer, $L_{x,\infty}$, and the surface $T_{s\infty}$, the temperature measured by a distant observer depend on the surface temperature, the neutron star radius and the surface redshift.

The surface temperature depends on the cooling of the inner neutron star by neutrino emission and also on the composition of the crust boundary layer. 
According to  \citep {Yakovlev+Pethick04, Potekhin+20},   $T_{s,\infty}$. starts at early times with 0.2 KeV and dropping to about 0.15 KeV after $ \sim 10^4$yr. The value of $L_{x,\infty}$, the luminosity observed by a distant observer starts at $2.5\times 10^{34} erg s^{-1}$ and after $   10^4$yr deceases to $10^{34} er s^{-1}$.  
For an  X-ray luminosity of $ 10^{34}\ {\rm erg\   s^{-1}}$  the flux from each galaxy is 
   
  \begin{equation}
  F_x= 5\times 10^8     10^{34} ( 4 \pi d^2) ^{-1}=   2.3 \times 10^{-11}{\rm erg\   s^{-1} cm^{-2}}.
  \end{equation}
and for the 100 galaxies at that distance the flux is  $ 2.3 \times 10^{-9}{\rm erg\   s^{-1} cm^{-2}}$,  three orders of magnitude larger than the observational value of  \cite{Cappelluti+17} referred to in section 3.1,

   \section{A "sharp"  transition lasting  $10^4$ yr}  
   
The papers proposing the late G transition refer to it as a sharp one. We dealt with the implications of  such a   sharp transition and found that it contradicts observational data of background neutrino and X-ray fluxes.

One may wonder about the viability of a transition lasting for a longer time which however is still {\it very short} compared to the look back time of $1.36 \times 10^8$ yr.  
 As an illustrative example we examine a transition  duration of $ 10^4\ yr$ corresponding to $\Delta z \simeq 10^{-6}$.   In this case, the average rate of energy deposition, as viewed by an observer far away from the neutron star  is

\begin{equation}
 L_{\infty}=7\times 10^{52}/ (3.15 \times 10^{11}) erg/s= 2.22 \times 10^{41}erg/s					
 		\end{equation}

 In a steady state, the outgoing   luminosity   $L_{\infty}$  equals the sum of the volume energy luminosity  of neutrinos and the surface luminosity of photons.  Following    \cite{Yakovlev+Pethick04}, the relevant equations, in this case, are:
 
\begin{eqnarray}
L_{\infty}= \int_0^R Q_{\nu} e^{2\phi(r)}e^{\lambda(r)}4\pi r^2 dr + \sigma_{sb} \frac{4  \pi R^2T_s^4}{ (1+z_s)^2}\\
T(r) e^{\phi(r)}= T_c e^ {\phi(0)}
\end{eqnarray} 
where $Q_{\nu}$ is the neutrino emissivity per unit volume,  
 $\sigma_{sb}$ is the Stephan-Boltzmann constant, and $Tc= T(r=0)$
Equation (18) is due to the high heat conductivity in the neutron star interior which implies a constant red-shifted temperature. 

The outermost layer of the neutron star(the thermal boundary layer) occupies a depth of $\sim 100 \ m$.  \cite{Yakovlev+Pethick04}  and \cite {Beznogov+21} quote the relation, that was derived by \cite{Gudmundsson+83}, between the temperature at the inner radius of this layer, $T_b$ and the surface temperature, $T_s$, . A similar relation was obtained by \cite{Beznogov+21}.

\begin{equation}
T_b =   1.288\times 10^8 \left(\frac{T_{s6}}{g_{14}}\right)^{0.455}K \ , \  \  g_{14}= \frac{G M e^{-\phi(R)}}{R^2 10^{14} cm\  s^{-2}}
\end{equation}
where  $T_{s6}=\frac{T_s}{10^6 K}$.

  Above a critical density $Q_{\nu}$ is   the  sum of the direct and modified  Urca processes
 while for  lower densities only the Modified Urca process is enabled. In \cite{Yakovlev+2001}
the critical density is $ 1.3\times 10^{15}\ gr \ cm^{-3}$ and in \cite{Yakovlev+Pethick04} it is   $7.85 \times 10^{14}\ gr \ cm^{-3}$. We adopt   a  value of $1\times 10^{15}\ gr \ cm^{-3}$.
  
   In what follows we employ representative median values for non-superfluid  nucleon matter based on  \cite{Yakovlev+Pethick04} 
  \begin{eqnarray}
    Q_{\nu, D}(r)= 1.5 \times 10^{27}\ erg\ cm^{-3} s^{-1}T_9^6 (r) \\
 Q_{\nu, M}(r)= 1.5 \times 10^{21}\ erg\ cm^{-3} s^{-1}T_9^8 (r)  
   \end{eqnarray}
   where    $ Q_{\nu, D}$ and $ Q_{\nu, M}$ are the direct and modified Urca emissivities, respectively, and $T_9$ is the temperature in units  of $10^9$K.
For a given neutron star model, the solution of equations(17)-(21)  yields the value of $T_c$ which then determines the surface temperature  as well as  the neutrino and photon luminosities far from the neutron star. In Table 2.  we present the results for the pre-transition
models of Table 1. 
 
\begin{table}
 
	\centering
	\caption{$R_D$ in units of km is the radius within which the direct Urca process occurs,$T_{c, 9}$  is the central temperature in units of $10^9$ K, $T_{s, \infty}$ is the surface temperature at infinity in units of $10^6$ K, $L_{x, \infty}$ is the X-ray luminosity far from the neutron star in units of $ 10^{36} erg s^{-1}$}
         \begin{tabular}{c c c c c c c}
   $ M_b $ & $M$&R&$R_D$ & $T_{c, 9}$ &$T_{s,6,\infty}$&$L_{x, 36, \infty}$ \\
\hline
\hline
  1.60 & 1.41&10.8& &1.01&114&2.25$\times 10^{5}$\\ 
  1.77 &1.54 &10.9& &0.99&119&2.2$\times 10^{5}$\\ 
  2.06&1.75&10.97&2.9&0.29&8.34&7.9\\
  2.20 &1.86&10.92&4.6&0.24&5.73&1.85\\
   \hline                                                                                 
 \end{tabular}
 \end{table}                                                                                                                                                                                                                                                
In the first two models the central density is below the critical density for the onset of the direct Urca process  and only the modified Urca process acts in the entire star. In the last two models the central density is above the critical density and the direct Urca process  operates up to a radius $R_D$ where the density  falls to the critical value. The modified Urca process operates in the entire star. However, the total contribution of the direct process is a factor of $\sim 10^4$ larger than that of the modified process. Consequently, the temperature at the star center is lower and so is the surface temperature, and the photon luminosity, than that in the two first models. The neutrino  luminosity is much larger but the neutrino energy is even lower making them undetectable.
The X-ray photons will be harder than in the sudden transition and the luminosities will be higher. So both for the MW and for the external galaxies, the X-ray fluxes would  be orders of magnitude above the observational ones.

eight
The structure of a given neutron star model changes continuously from the pre-transition to the post-transition solution, of table 1.  The direct Urca process
will at some time turn on  for the second model while for  the two last models $R_D$ will increase and so will the effectiveness  of the direct Urca process. Thus, lower central and surface temperatures are expected as well as a lower X-ray luminosity far from the neutron star.

 In order to get an idea about the changes, we repeat the computations for the post-transition third model  in Table 1. ($ M_b=2.06 M_{\odot}$).
The results are: $R_D$ is now $5.66$ km, $T_{c, 9}= 0.21$, $T_{s, 6,\infty}= 4.7 $, and $L_{x,  \infty}= 3.5\times 10^{34}$. Therefore, even towards the completion of the transition, the X-ray luminosity at infinity is larger than in the sudden transition, and  so will be also the fluxes both from neutron stars in the MW galaxy and from  neutron stars in the external galaxies.

\section{Discussion and Conclusions}
   Due to their compactness
and large gravitational binding energy,  neutron stars are a natural test ground for the late G transition. To this adds the fact that {\it all }the neutron stars that  formed  at red shift $z>0.01$ will undergo a thermal revival by the G transition. Since there are about $5 \times 10^8$ such neutron stars in  a MW type galaxy the total thermal energy per galaxy will be $\sim 3.5\times 10^{61}$ ~erg. Most of the energy would be emitted via neutrinos.
 
 We considered two 
cases. The first one is an abrupt step function  G transition and the second involves a gradual one  over a representative time span of $10^4$~years that is still very short compared to the look-back time of $1.36 \times 10^8$ yr, corresponding to $z=0.01$.

In the first case, the energy flux  of $\sim 10$~MeV neutrinos from external galaxies, is a factor of $\sim 22$ larger than the observational upper limit. The X-ray flux from the MW galaxy exceeds by seven  orders of magnitude the observational upper limit on diffuse  unresolved X-ray background at energies of $0.5-1$~KeV. The X ray flux from external galaxies exceeds that limit by three orders of magnitude.

In the second case,   the neutrino energies will  be $\leq  0.1$~MeV and thus no observational limit exists. But the X-ray flux, from MW neutron stars and from external  galaxies will be even larger than  in the first case.

Can  X-ray absorption seriously decrease the expected fluxes? Using a cross section for absorption in the interstellar medium for 0.1 KeV X-rays of $10^{-28} cm^{-2}$ per Hydrogen atom \cite{Wilms+2000}, the optical depth for absorption in the warm neutral medium with hydrogen number density of $1 cm^{-3}$
 is   $6.2 \times 10^{-5} L/(20 Kpc)$ along a line of length $L$. Even for the cold neutral medium with number density of $1000 cm^{-3}$ one gets an optical depth of  $6.2 \times 10^{-2} L/(20 Kpc)$. So absorption is not important even for  X-rays from the MW neutron stars and from  edge-on external galaxies.

Therefore, one may conclude that the proposed late  G transition  scenario    is non-viable.
\section*{Acknowledgements}
I thank Shmuel Nussinov for interesting discussions.

\section*{Declaration of competing interests} 
The author declares that he  has no known competing financial interests or personal relationships that could have appeared to influence the work reported in this paper.

\section*{Data Availability}
No data sets have been used in this paper; only citations of observational results.

   \end{document}